# Improving Energy Management of Hybrid Electric Vehicles by Considering Battery Electric-Thermal Model

A. Mousaei
Dept. Electrical and Computer Engineering
University of Tabriz
Tabriz, Iran
a.mousaei@tabrizu.ac.ir

*Abstract*— **This article proposes an offline Energy Management System (EMS) for Parallel Hybrid Electric Vehicles (PHEVs). Dividing the torque between the Electric Motor (EM) and the Internal Combustion Engine (ICE) requires a suitable EMS. Batteries are vital to HEVs and significantly impact overall vehicle cost and performance. High temperature and high battery State of Charge (SOC) are the main factors that accelerate battery aging. SOC is the most critical state variable in EMS and was usually considered the only dynamic variable in previous studies. For simplicity, the battery temperature was often assumed to be constant, and the effect of EMS on temperature change was neglected. In this paper, we first apply Dynamic Programming (DP) to a PHEV without considering battery temperature variations. Then, the battery model is improved by modeling the cooling system to take into account temperature variations and show how neglecting the thermal dynamics of the battery in EMS is impractical. Finally, by integrating battery temperature as a state variable in the optimization problem, a new EMS is proposed to control battery temperature and SOC variation. Simulation results of the tested vehicle show that the proposed method controls battery charge and temperature. The proposed EMS method prevents uncontrolled fluctuations in battery temperature and reduces its deterioration rate.**

I. INTRODUCTION

Two-thirds of the petroleum used worldwide is consumed by cars, and about half of this amount is related to passenger cars. Pollution caused by fuel consumption and dependence on external sources has motivated much research and advances in replacing conventional power generation and transmission systems based on Internal Combustion Engines (ICE) with renewable and clean energy sources. On the other hand, the power of pure Electric Vehicles (EVs), due to the high cost and low capacity of the batteries, does not meet the general needs except for special applications. As a result, one of the primary motivations for the production of Hybrid Electric Vehicles (HEVs) has been to take advantage of the high power of Conventional Vehicles (CVs) and the low emission of EVs [1].

HEVs have two directions to provide their driving force: ICE, the Electric Machine (EM), and the battery. The most crucial issue for control engineers in these vehicles is the issue of EMS. The proper performance of HEVs, regardless of the type of structure and the characteristics of various components, is highly dependent on the EMS. The control algorithm of EMS in HEVs specifies how to divide the demand power or demand torque between the electric and thermal units of the vehicles to follow a driving cycle according to the minimal fuel consumption and emission of pollutants [1,2].

In the last two decades, various methods have been presented for the EMS of these vehicles [1-7]. The control theory of EMSs, divide into two categories: based on rules and based on optimal control. Although rules-based methods are more straightforward, they do not provide the optimal response, and optimal control-based methods are used to obtain the optimal response.

The standard classification of methods based on optimal control divides them into offline and online forms. All offline methods' common assumption knows all road and driving conditions. This assumption is necessary to reach the optimal answer. Although the offline methods are not directly applicable to the online control of vehicles, their response can be used to select the best performance, compare the online techniques, and define reference routes in online mode. In online forms, a cost function is usually minimized. Therefore, all online methods are suboptimal, and their measurement criteria are close to the overall optimal results obtained by offline methods. Online methods that only use momently information have far from the optimal answer. The methods that use pattern recognition and processing previously stored information depend highly on the conditions and the driving cycle for which they are designed. They do not have a suitable answer for other driving cycles [1-8]. The essential methods based on optimal control are Dynamic Programming (DP) for offline energy management [3,9] and Pontriagin's Minimum Principle (PMP) for online energy management [4,9].

In the standard EMS, the cost function is the amount of fuel consumption, and the only dynamic variable is the State of Charge (SOC) of the battery. Therefore, parts of the vehicle, like the battery, are modeled using lookup tables and static maps. Due to the size and cost of the batteries, especially in plug-in HEVs, the vehicle's battery is very vulnerable to high temperatures. So, their continuous charging and discharging and optimal battery use are necessary and vital [5].

Many studies have been done to optimize the use of batteries in HEVs. [8] and [10,11] have studied the modeling of battery temperature changes. A group of articles has modeled battery temperature changes without including them in the EMS problem. Another group of articles also considered the State of Health (SOH) of the battery in the optimization problem but assumed the battery temperature to be constant. In [12-14], battery life and its degree of SOH are modeled. Unsuitable working conditions reduce battery life and cause increased internal resistance and a decrease in capacity, and [12] and [14] have modeled this action. In [7], the battery life is considered in the EMS problem, but the battery temperature is assumed to be constant. The design of different systems for the thermal management of batteries has been studied in [15] and [16]. References [17] and [18] have presented methods to estimate battery SOC, and [19] the optimal range of battery temperature has been investigated. In [20], the battery's lifetime depends only on its current. Still, a more accurate model must consider other parameters affecting the battery's performance for a general conclusion. In [21], it was viewed as an equation of state instead of evaluating the SOH in the cost function. Because the ranges of SOH and SOC variables are very different, two controllers were used to follow the reference, but the optimality of the response was not guaranteed. Considering that battery temperature is a critical and essential factor in corrosion, a penalty for battery temperature was considered [22]. But battery temperature is the only factor regarded in this reference, while other elements effectively reduce battery lifetime, even at mild temperatures. In [23], optimal energy management was presented with a cost function that includes fuel consumption and battery health. This article used a model for corrosion dependent on battery temperature and the current passing through it. The limitation of this article was that the vehicle's battery was modeled statically, and the battery's temperature was assumed to be constant. [24] has shown that the battery lifetime of vehicles in temperateness (on average between -3 °C to 32 °C) is 73 to 94% more than in tropic (up to 32 °C) areas. The reference of [25] presented a method to estimate battery lifetime and investigated the effect of driving style on the battery in addition to temperature and current. The reviewed references show the harmful effects of increasing the battery temperature on the battery and the vehicle's performance. Still, it has not been presented in the form of EMS. Also, the references that have addressed the issue of EMS have considered the battery temperature as constant or ignored the dependence of the parameters on the battery temperature.

In this article, an offline EMS using DP is first implemented for an HEV whose model is presented in [1]. In this model of the HEV, a cooling system that uses air as a heat transfer fluid has been used. The battery model is improved by modeling this cooling system, including the battery's electric and thermal dynamics. With this, battery temperature changes are modeled as a dynamic variable, and then battery temperature is added to the fuel consumption optimization system as a controlled state variable. This way, the proposed energy management system will have two equations of state: SOC and battery temperature. It will be shown that battery temperature and SOC are under control during the energy management process and do not go out of the desired range.

Therefore, the innovations of this article are:

1. Improving the vehicle model by modeling the battery cooling system and considering battery temperature changes for an HEV.

2. Offline energy management by DP method with two equations of the state, including SOC and battery temperature, in the optimization problem based on the improved model.

## II. DYNAMIC MODEL OF HEV

Unlike conventional and electric vehicles, HEVs have at least two sources of power to propulsion (figure1): An ICE or a fuel cell as a fuel converter and an Electric Machine (EM) with a battery, supercapacitor, or flywheel as a source of energy storage.

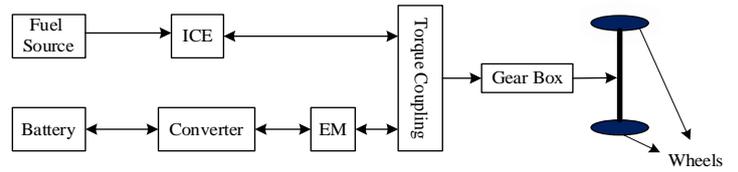

Fig. 1. The structure of the Parallel HEV

For energy management, the longitudinal dynamics of the vehicle are used to model the vehicle's chassis, which calculates the car's speed according to equation (1) with the thrust force. Figure 2 shows the relation (1).

$$m\frac{dV(t)}{dt}=F_t(t)-[F_a(t)+F_r(t)+F_g(t)+F_d(t)] \quad (1)$$

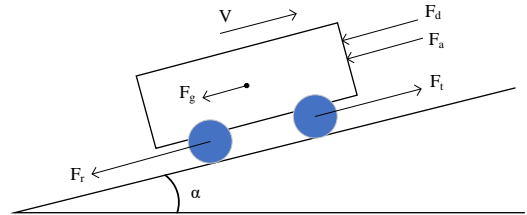

Fig. 2. Forces acting on the vehicle

## III. MODEL OF BATTERY

The battery of HEVs is formed by connecting several identical cells in series. Therefore, one cell is usually modeled, and the battery's output voltage will be the cells' total voltage. In the discussion of energy management, the electrical equivalent circuit of Figure (3) is usually used to model the battery. This equivalent circuit includes the internal resistance of the battery ($R_b$) and an ideal voltage source called $V_{oc}$ as the battery's open-circuit voltage. For simplicity, the battery temperature is assumed to be constant, and the internal resistance and open-circuit voltage will depend only on the SOC.

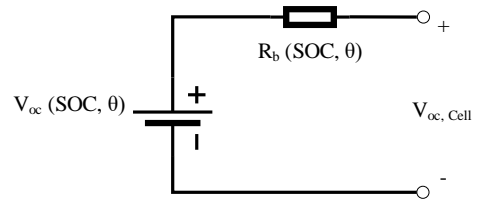

Fig. 3. Equivalent circuit of the battery

According to equation (2), the SOC of the battery will be calculated. We have:

$$SOC(t) = \frac{S(t)}{S_0} \quad (2)$$

As mentioned in the introduction, to control the battery's temperature, increase its lifetime, and improve the vehicle's performance, it is necessary to consider the changes in the temperature of the battery in the EMS. For this purpose, the proposed model in [12] and [14] is used for the battery. The presented model for the battery consists of two electrical and thermal parts. The electrical model calculates the battery voltage and SOC, and the thermal model predicts the battery temperature. The electrical sub-system of the battery model is shown in Figure (3). For this model, SOC changes can be calculated as follows:

$$\frac{dSOC(t)}{dt} = -\frac{I_b(t)}{S_0} \qquad (3)$$

$$V_o(t) = V_{oc}(t) - R_b(t)I_b(t) \qquad (4)$$

In the above relationships, S(t) is the battery's current charge, $S_0$ is its total capacity, and $I_b$ is the current of the battery, which is obtained by equation (5).

$$I_b = \frac{V_{oc} - \sqrt{V_{oc}^2 - 4R_b P_m}}{2R_b} \qquad (5)$$

To examine the thermal model of the battery, its cooling system is shown in Figure (4). This system uses air as a heat transfer fluid. This system is less complex than the system that uses liquid. Using air as a heat transfer fluid system works well for parallel HEVs, but a liquid cooling system is preferred for series HEVs. Figure (4) shows the temperature of the air entering the channel. According to passing air through the channels, the battery is cooled and the temperature of the air at the exit of the channels reaches $\theta_{out}$.

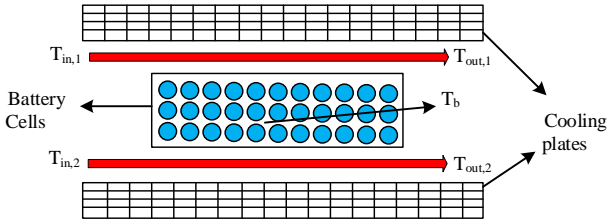

Fig. 4. Cooling system with two channels

In the thermal model of the battery, based on the law of conservation of energy, the temperature changes of the battery cell follow equation (6), where $\theta$ is the battery temperature, $Q_g$ is the speed of the heat generation, and $Q_d$ is the speed of depletion heat by the channels of the cooling system. Also, $m_C$ and $C_{P,C}$, are the battery cell's weight and specific heat capacity, respectively.

$$m_C C_{P,C} \frac{d\theta}{dt} = Q_g - Q_d \qquad (6)$$

In [28], the speed of the heat generation is shown by equation (7). Equation (7) $\frac{\partial V_{oc}}{\partial \theta}$, can be ignored compared to other sentences [11].

$$Q_g = I_b \cdot (V_{oc} - V_o - \frac{\partial V_{oc}}{\partial \theta}\theta) \qquad (7)$$

The heat removed by the cooling system, according to equation (8), includes the heat removed from channels 1 and 2.

$$Q_d = Q_{ku,1} + Q_{ku,2} \qquad (8)$$

When the air enters the channels with a temperature lower than the temperature of the battery, the heat of the battery is removed through surface convection by the channels. The speed of heat removal from each channel is calculated by equation (9). According to this equation, the heat exchange speed in the channels' output is a function of the temperature at the channel's output and the temperature of the battery cell.

$$Q_{ku,i} = \bar{h}_i A_{ch}(\theta - \theta_{out,i}) \qquad (9)$$

Where $A_{ch}$ is the area of the heat transfer of the channel, $\bar{h}_i$ is the average heat transfer coefficient, and $\theta_{out,i}$ is the air temperature at the outlet of channel i. According to the energy balance equation, the heat exchange speed at the channels' entrance is equal to the speed of heat exchange at the exit of the channels. Therefore, relation (10) can be written.

$$Q_{ku,i} - Q_{u,i} = 0 \qquad (10)$$

The speed of the heat exchange at the entrance of the channels, which is a function of the temperature of the air entering and exiting the channel, is obtained by equation (11).

$$Q_{u,i} = \rho_{air} C_{p,air} q_{p,air} (\theta_{out,i} - \theta_{in,i}) \qquad (11)$$

That, the $\rho_{air}$ is the air density, $C_{p,air}$ is the specific heat capacity, and $q_{p,air}$, is the volume flow speed of air in the channels. Using relations (8) to (11) and removing $\theta_{out,i}$ from these relations, equations (12) to (17) are obtained to calculate the heat dissipation rate ($Q_d$).

$$Q_d = \alpha_1 \theta + \alpha_2 \theta_{in,1} + \alpha_3 \theta_{in,2} \qquad (12)$$

$$\alpha_1 = \frac{1}{R_{ku,1} + R_{u,1}} + \frac{1}{R_{ku,2}} \qquad (13)$$

$$\alpha_2 = -\frac{1}{R_{ku,1} + R_{u,1}} \qquad (14)$$

$$\alpha_3 = -\frac{1}{R_{ku,2} + R_{u,2}} \qquad (15)$$

$$R_{ku,i} = \frac{1}{\bar{h}_i A_{ch}} \qquad (16)$$

$$R_{u,i} = \frac{1}{\rho_{air} C_{p,air} q_{p,air,i}} \qquad (17)$$

According to equations (6) to (17), the equation of state for battery temperature will be according to relation (18).

$$\dot{\theta} = \frac{R_b I_b^2}{m_C C_{P,c}} - \frac{\alpha_1 \theta + \alpha_2 \theta_{in,1} + \alpha_3 \theta_{in,2}}{m_C C_{P,c}} \qquad (18)$$

## IV. ENERGY MANAGEMENT METHOD

Optimal energy management consists of finding the control law to minimize the cost function during the driving cycle considering the driving needs and the various constraints defined on the state and control variables. There are many expressions for optimal control [1]. The cost function in energy management methods includes criteria such as fuel consumption, pollutant emissions, drivability, gradeability, or a combination [32]. The proposed method selects the cost function according to equation (19). In this cost, the function $\dot{m}_f$ is fuel consumption speed.

$$J = \int_0^{T_f} \dot{m}_f(\omega(t), u(t)) dt \qquad (19)$$

The control variable u in this system is the ratio of torque division. Having the total required torque in the wheels of the vehicle, $T_w$, and the ratio u, which the supervisory controller determines, according to the relations (20) and (21), the demanded torque in the ICE and the EM, $T_e$, and $T_m$, respectively can be calculated.

$$T_m(t) + T_e(t) = T_w(t) \qquad (20)$$

$$u(t) = T_m(t)/T_w(t) \qquad (21)$$

According to the above contents and relationships (1) to (21), the vehicle's final model combines dynamic equations and maps. The calculation process in this model is that first, the traction force ($F_t$) and the required torque of the vehicle ($T_w$) to reach the driving cycle speed (v) are obtained. Then, according to the u, the torque of the EM and ICE is calculated. In the next step, having the speed and torque of the motors, their power is obtained with the help of efficiency maps of these motors. The following relationship (equations (22) to (27)) is related to the battery current. Finally, the changes in the state of charge and temperature of the battery are calculated as state variables with the last two relationships of this model. In the standard energy management method used in most previous references, the only dynamic state variable of the system is the SOC of the battery, and the static model is used for the rest of the vehicle components.

Using the reference tables for the battery model and ignoring the thermal dynamics and temperature changes of the battery will cause damage to the battery and reduce its lifetime. Since the battery of HEVs is the most sensitive and expensive component, its performance dramatically impacts the vehicle's performance. It is necessary to keep SOC and its temperature within a specific range for optimal vehicle performance. In this article, to control the temperature of the battery in addition to its SOC during energy management, taking into account the electric-thermal dynamics of the battery, the energy management problem is defined with two state variables, SOC and θ with relations (22) to (27).

$$x(t)=[SOC(t) \quad \theta(t)]^T \tag{22}$$

$$\dot{x}(t)=\begin{bmatrix} -\dfrac{I_b}{S_0} \\ \dfrac{R_b I_b^2}{m_c C_{P,c}} - \dfrac{\alpha_1\theta+\alpha_2\theta_{in,1}+\alpha_3\theta_{in,2}}{m_c C_{P,c}} \end{bmatrix} \tag{23}$$

$$SOC_{low} \leq SOC(t) \leq SOC_{high} \tag{24}$$

$$SOC(T_f) \in [SOC_{N,min}, SOC_{N,max}] \tag{25}$$

$$\theta_{low} \leq \theta(t) \leq \theta_{high} \tag{26}$$

$$\theta(T_f) \in [\theta_{N,min}, \theta_{N,max}] \tag{27}$$

Methods such as Linear Programming (LP), Dynamic Programming (DP), and evolutionary algorithms are used to find the optimal answer to the energy management problem of HEVs. DP has the most use and the best performance among all the implemented methods because it guarantees the ultimate optimal solution based on Bellman's optimality principle [33] and [34]. As will be shown below, the problem optimal energy management (equations (22) to (27)) is fully compatible with the form of the problem in DP (equations (28) to (33)).

DP is a very suitable method for finding the optimal control and path of state variables in problems of the form (26) problems in which K= 0, 1,…, N.

In relations (28) to (33), $g_N(x_N)$ is a penalty to limit the final value of the state variable, and $g_k(x^k, u_k)$, is the Cost-to-go function. The value of this function is the cost of applying $u_k$ at the moment k to a system with state function $g_k(x^k, u_k)$ and initial value $x_0$.

$$\min_{u_k \in U_k} \{g_N(x_N) + \sum_{0}^{N-1} g_N(x_k, u_k) \tag{28}$$

$$x_{k+1}=f_k(x_k, u_k) \tag{29}$$

$$x_0=x0 \tag{30}$$

$$x_N \in T \subset R^n \tag{31}$$

$$x_N \in X_k \subset R^n \tag{32}$$

$$u_N \in U_k \subset R^n \tag{33}$$

Since DP is a numerical method, the continuous control problem should be discretized from (28) to (33). The method's accuracy depends on the meshing accuracy or the resolution of time and state spaces.

Another critical issue is the value of the cost function for impossible states. Allocation of infinite transfer cost is the first method proposed for such points. As said, interpolation is used to determine the cost function of transmitting the middle points of space networking. Then doing interpolation for points adjacent to impossible states will cause those points to be interpreted as unbelievable points, which is the fundamental problem of assigning the infinite value. Some methods were proposed to solve this problem; among them, it is possible to mention giving a huge value instead of an infinite value [36].

The DP algorithm calculates the optimal cost function $J_k(x^i)$ as follows, starting from the end of the path (k=N) and performing recursive calculations until k=0 and for all points of the discrete time-state space:

1. *Determining the initial value of the cost function in k=N.*

$$j_N(x^i)=\begin{cases} j_N(x^i) & x^i \in T \\ \infty & else \end{cases} \tag{34}$$

The transfer cost function is zero at the last moment because there is no next state. In problems where the final value of the state variable is bound, $j_N(x^i)$ is equal to the penalty for the deviation of the final value. According to (19), the final value of the SOC and battery temperature variables is limited in the energy management problem. This limitation is determined by the set T in (31). At this stage, according to the final values of SOC and θ outside the limits defined in the problem, the value ∞ (infinite) is assigned. So, the paths leading to these values will not be the optimal path and the answer to the problem.

2. *Calculation of the optimal cost function $j_N(x^i)$ at all points of the discrete space for k=N-1 to k=0 and for $x^i \in X_k$.*

$$j_N(x^i)=\min_{u_k \in U_k}\{g_k(x^i, u_k)+j_{k+1}(f_k(x^i, u_k))\} \tag{35}$$

In the intermediate steps, an optimal path is found for each state space point until the path's end and stored for use in the following steps. Each sub-problem is solved once and its answer is saved, and with this, the repetition of sub-problems is prevented when their answers are needed again. The transfer cost function $j_N(x^i)$ is the cost of moving from the point $x^i$ at the moment k to the end of the optimal path, which is formed by two terms $g_k(x^i, u_k)$ and $j_{k+1}(f_k(x^i, u_k))$. The first term is the cost of the transfer from the point $x^i$ at the moment k to the point $f_k(x^i, u_k)$ at the moment k+1, and the second term is the cost of the optimal path from this new point to the end of the path, which was calculated and stored in the previous step. This way, by continuing this algorithm until k=0, the optimal control signal is obtained at every moment. Finally, the optimal control sequence $\pi = \{\mu_0, \mu_1, ..., \mu_{N-1}\}$ is obtained.

## V. SIMULATION RESULTS

The vehicle under investigation is the Hyper Daimler Chrysler parallel Electric Hybrid Vehicle from the Mercedes A-Class series, presented in [1]. The values of the general parameters of the vehicle are given in Table I, and the specifications of the vehicle's battery are shown in Table II.

TABLE I. GENERAL PARAMETERS OF THE VEHICLE

| Parameter | Value (Unit) |
|---|---|
| Vehicle weight | 1800 (Kg) |
| Wheel radius | 0.3 (m) |
| Rolling friction | 144 (N) |
| Aerodynamic coefficient | 0.48 (N.s$^2$.m$^{-2}$) |
| The maximum torque of the ICE | 199 (N.m) |
| The maximum torque of the EM | 133 (N.m) |
| The maximum speed of the ICE | 503 (rad.s$^{-1}$) |
| The maximum speed of the EM | 600 (rad.s$^{-1}$) |

TABLE II. LI-ION BATTERY MODEL PARAMETER VALUES

| Parameter | Value (Unit) |
|---|---|
| Length | 0.19 (m) |
| Width | 0.145 (m) |
| Thickness | 0.005 (m) |
| Mass | 3.84 (Kg) |
| Nominal capacity | 15 (A.h) |
| Nominal voltage | 3.75 (V) |
| Specific heat capacity | 800 (J.kg$^{-1}$.K$^{-1}$) |

The driving cycle used in the simulation is the JN-1015 (Japan), with the specifications listed in Table III. Also, the speed-time diagram of JN-1015 is shown in figure 5.

TABLE III. SPECIFICATIONS OF JN-1015 (JAPAN)

| Parameter | Value (Unit) |
|---|---|
| Distance | 4165.27 (m) |
| Time | 660 (s) |
| The highest speed | 19.44 (m.s$^{-1}$) |
| Average speed | 8.5 (m.s$^{-1}$) |
| Average acceleration | 0 (m.s$^{-2}$) |

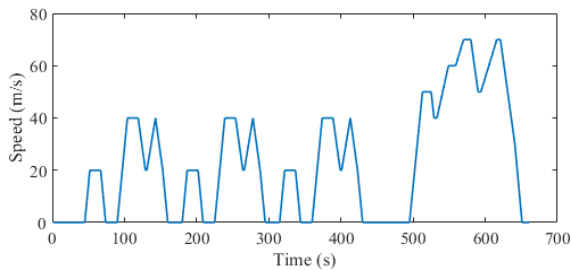

Fig. 5. Driving cycle JN-1015 (Japan)

First, by ignoring the effect of the thermal dynamics of the battery, the changing temperature is not considered in the relationships of the optimization problem. The only variable is the SOC, and the values of its parameters are given in Table IV.

TABLE IV. VARIABLES IN THE SOC

| Parameter | Value |
|---|---|
| SOC$_{low}$ | 0.4 |
| SOC$_{high}$ | 0.7 |
| SOC$_0$ | 0.5 |
| SOC$_{N, min}$ | 0.54 |
| SOC$_{N, max}$ | 0.55 |

Fuel consumption in this method is equal to 4.3 liters per 100 kilometers. Figure 6 shows SOC changes. As mentioned, in HEV that the network cannot recharge; the final charging level should be close to its initial value. It can be seen in Figure 6 that SOC approaches the initial value at the end of the path.

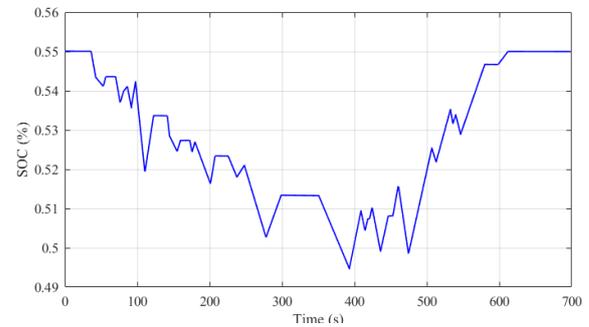

Fig. 6. SOC for energy management without battery temperature control

Figure 7 shows how to divide the demanded torque between EM and ICE by the control variable $u(t)=T_m(t)/T_w(t)$. Due to the power of the EM being less than the ICE, high torques are inevitably provided by the ICE. For amounts of torque that both EM and ICE can provide, the priority is with the EM, provided that the battery charge level does not fall below the minimum allowed charge. Negative torques are also used by the EM (in generator mode) to charge the battery.

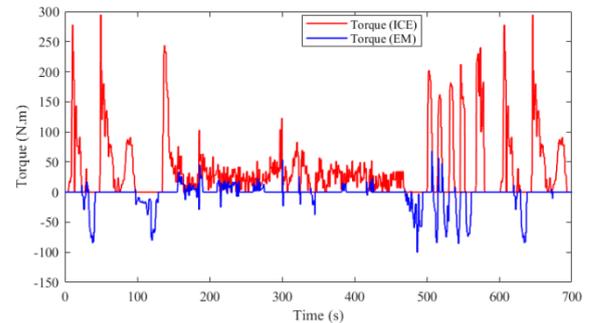

Fig. 7. Torque split between EM and ICE

As mentioned in the optimization problem in this section, the battery's temperature was considered constant. Now, using the parameters of the battery model and the resulting internal resistance and battery current vectors, we will show the effect of the EMS on the battery temperature.

In figure 8, it can be seen that the battery's temperature has increased to about 57.2 °C due to being neglected in the optimization despite the presence of the cooling system. In the following, the results of the proposed method are shown. In this method, the improved battery model described in section 4.2 is used in the optimization problem. Battery temperature is considered the second state variable next to SOC in dynamic planning. SOC parameter values are the same as in table IV, and battery temperature is set in table V.

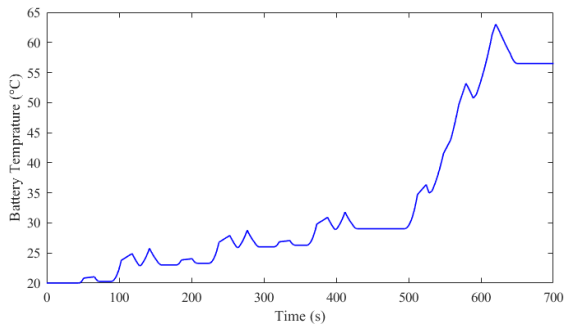

Fig. 8. Changing the battery temperature without control

TABLE V. SPECIFICATIONS OF JN-1015 (JAPAN)

| Parameter | Value (Unit) |
|---|---|
| $\theta_{low}$ | 10 (K) |
| $\theta_{high}$ | 30 (K) |
| $\theta_0$ | 20 (K) |
| $\theta_{N, min}$ | 15 (K) |
| $\theta_{N, max}$ | 25 (K) |

Adding upper and lower limits of the battery's temperature to the optimization problem has resulted in less battery usage. In this case, if providing the demanded torque causes the temperature to increase and approaches the upper limit of the temperature, it will be prevented. This will improve the operation of the ICE and, as a result, increase fuel consumption. Fuel consumption in this mode has reached 5.1 liters per 100 kilometers. Figures 9 and 10 show the proposed method's SOC and battery temperature changes.

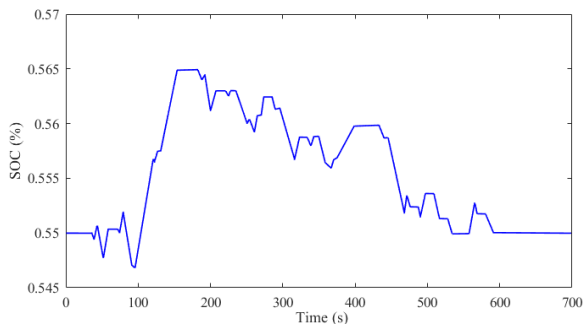

Fig. 9. SOC in the proposed method

It can be seen that in addition to SOC, the battery temperature also remains within the specified range, and its final value has reached 25.45 °C.

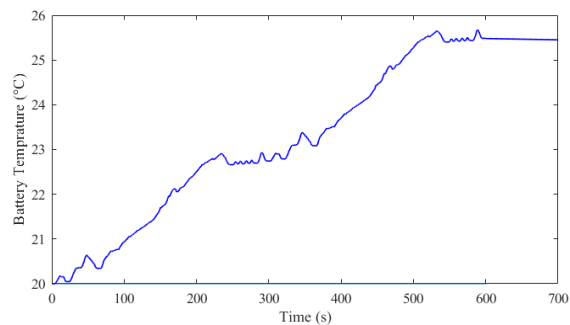

Fig. 10. Changing the battery temperature in the proposed method

## VI. ONCLOTION

This article proposed a method for offline energy management of Hybrid Electric Vehicles based on the optimal use of the battery for HEVs. The problem of optimal energy management of these vehicles is a non-linear problem with different performance limitations according to modeling and system variables. Dynamic Programming, a numerical solution method, is suitable for finding the optimal answer to such a problem offline. This method was applied to the vehicle under study with the assumption of constant battery temperature. By modeling battery temperature changes, it was shown that the assumption of continuous battery temperature is impractical, and the effect of energy management on battery temperature cannot be ignored. According to the results, by ignoring the control of temperature changes in the energy management strategy, the value of SOC from the initial value of 55% decreased to 49.4% by a change of 10.19%, then returned to 55%. Also, in this case, the battery's temperature reached 57.2 °C from 20 °C with a sharp change of 186%, which will cause many disadvantages, including correction and reduced battery life. In the proposed method, according to the control of the battery temperature changes, the value of SOC reached 56.6% from 55%, with an increase of 2.73%. Also, in the proposed strategy, the battery's temperature increased by 27.25% and reached 25.45 °C from 20 °C. According to this article's results, changes in SOC improved by 12.92% in the proposed method. Also, compared to conventional energy management methods, the proposed method reduced changes in the battery temperature by 6.82 times and increased battery life.

REFERENCES

[1] L. Guzzella and A. Sciarretta, Vehicle Propulsion Systems, Introduction to Modeling and Optimization, Springer, 2013.
[2] A. A. Malikopoulos, "Supervisory Power Management Control Algorithms for Hybrid Electric Vehicles: A Survey," IEEE Transactions on Intelligent Transportation Systems, vol. 15, no. 5, pp. 1869-1885, 2014.
[3] A. Mousaei, M. B. Bannae Sharifian and N. Rostami, "An Improved Fuzzy Logic Control Strategy of an Optimized Linear Induction Motor Using Super Twisting Sliding Mode Controller," *2022 13th Power Electronics, Drive Systems, and Technologies Conference (PEDSTC)*, 2022, pp. 1-5, doi: 10.1109/PEDSTC53976.2022.9767465.
[4] K. Namwook, C. Sukwonand P. Huei, "Optimal Control of Hybrid Electric Vehicles Based on Pontryagin's Minimum Principle," IEEE Transactions on Control Systems Technology, vol. 19, no. 5,pp. 1279-1287, 2011.
[5] A. Mousaei and M. B. B. Sharifian, "Design and optimization of a linear induction motor with hybrid secondary for textile applications," *2020 28th Iranian Conference on Electrical Engineering (ICEE)*,

[6] L. Serrao et al., "Open Issues in Supervisory Control of Hybrid Electric Vehicles: A Unified Approach Using Optimal Control Methods," Oil & Gas Science and Technology–Revue d'IFP Energies nouvelles, vol. 68, no. 1, pp. 23-33, 2013.
[7] G. Paganelli et al., "Equivalent Consumption Minimization Strategy for Parallel Hybrid Powertrains," in IEEE 55th Vehicular Technology Conference, VTC Spring 2002.
[8] A. Mousaei, M. B. Bannae Sharifian and N. Rostami, "Direct Thrust Force Control (DTFC) of Optimized Linear Induction Motor with Super Twisting Sliding Mode Controller (STSMC)," *2021 12th Power Electronics, Drive Systems, and Technologies Conference (PEDSTC)*, 2021, pp. 1-5, doi: 10.1109/PEDSTC52094.2021.9405903.
[9] L. Serrao et al., "Optimal Energy Management of Hybrid Electric Vehicles Including Battery Aging," in American Control Conference (ACC), IEEE 2011.
[10] C. H. Zheng et al., "The Effect of Battery Temperature on Total Fuel Consumption of Fuel Cell Hybrid Vehicles," International Journal of Hydrogen Energy, vol. 38, no. 13, pp. 5192-5200, 2013.
[11] D.E. Kirk, Optimal Control Theory: An Introduction, Courier Corporation, 2012.
[12] A. Mousaei, M. B. Bannae Sharifian and N. Rostami, "An Improved Predictive Current Control Strategy of Linear Induction Motor Based on Ultra-Local Model and Extended State Observer," *2022 13th Power Electronics, Drive Systems, and Technologies Conference (PEDSTC)*, 2022, pp. 12-18, doi: 10.1109/PEDSTC53976.2022.9767535.
[13] A. Cordoba-Arenas et al., "Capacity and Power Fade Cycle-Life Model for Plug-in Hybrid Electric Vehicle Lithium-ion Battery Cells Containing Blended Spinel and Layered-Oxide Positive Electrodes," Journal of Power Sources, vol. 278, pp. 473-483, 2015.
[14] R. Mahamud and C. Park, "Reciprocating Air Flow for Li-ion Battery Thermal Management to Improve Temperature Uniformity," Journal of Power Sources, vol. 196, no. 13, pp. 5685-5696, 2011.
[15] H. Park, , "A Design of Air Flow Configuration for Cooling Lithium-ion Battery in Hybrid Electric Vehicles," Journal Of Power Sources, vol. 239 (Supplement C), pp. 30-36, 2013.
[16] D. Di Domenico, E. Prada and Y. Creff, "An Adaptive Strategy for Li-ion Battery Internal State Estimation," Control Engineering Practice, vol. 21, no. 12, pp. 1851-1859, 2013.
[17] J. Kalawoun et al., "From a Novel Classification of the Battery State of Charge Estimators Toward a Conception of an Ideal One," Journal of Power Sources, vol. 279, pp. 694-706, 2015.
[18] J. Sun et al., "LiFePO4 Optimal Operation Temperature Range Analysis for EV/HEV," International Conference on Life System Modeling and Simulation and International Conference on Intelligent Computing for Sustainable Energy and Environment, pp. 476-485, Springer, Berlin, Heidelberg, 2014.
[19] L. Serrao et al., Optimal Energy Management of Hybrid Electric Vehicles Including Battery Aging. In: Proceedings Of The IEEE American Control Conference (ACC), San Francisco, Ca, Jun 29–Jul 1, 2011.
[20] S. Ebbesen, P. Elbert, and L. Guzzella, "Battery State-of-Health Perceptive Energy Management for Hybrid Electric Vehicles," IEEE Transactions on Vehicular Technology, vol. 61, no. 7, pp. 2893–2900, Sep., 2012.
[21] T. M. Padovani et al., "Optimal Energy Management Strategy Including Battery Health Through Thermal Management for Hybrid Vehicles," IFAC Proceedings Volumes, vol. 46, no. 21, pp. 384-389, 2013.
[22] L. Tang, G. Rizzoni, and S. Onori, "Energy Management Strategy for HEVs Including Battery Life Optimization," IEEE Transactions on Transportation Electrification, vol. 1, no. 3, October 2016.
[23] T. Yuksel et al., "Plug-in Hybrid Electric Vehicle LiFePO4 Battery Life Implications of Thermal Management, Driving Conditions, and Regional Climate." Journal of Power Sources, vol. 338, pp. 49-64, 2017.
[24] M. Jafari et al., "Electric Vehicle Battery Cycle Aging Evaluation in Real-World Daily Driving and Vehicle-To-Grid Services." IEEE Transactions on Transportation Electrification, vol. 4, no. 1, pp. 122-134, 2018.
[25] T. Nüesch et al., "Equivalent Consumption Minimization Strategy for the Control of Real Driving NOx Emissions of a Diesel Hybrid Electric Vehicle," Energies, vol. 7, no. 5, pp. 3148-3178, 2014.
[26] A. Pesaran, M. Keyser and S. Burch, "An Approach for Designing Thermal Management Systems for Electric and Hybrid Vehicle Battery Packs," National Renewable Energy Laboratory, Golden, CO (US), 1999.
[27] D. Bernardi, E. Pawlikowski and J. Newman, "A General Energy Balance for Battery Systems," Journal Of the Electrochemical Society, vol. 132, no. 1, pp. 5-12, 1985.
[28] K. B. Wipke et al., "ADVISOR 2.1: A User-Friendly Advanced Powertrain Simulation Using a Combined Backward/Forward Approach." IEEE transactions on vehicular technology, vol. 48, no. 6, pp. 1751-1761, 1999.
[29] T. Markel et al., " ADVISOR: a Systems Analysis Tool for Advanced Vehicle Modeling." Journal of power sources, vol. 110, no. 2, pp. 255-266, 2002.
[30] T. Barlow et al., "A Reference Book of Driving Cycles for Use in The Measurement of Road Vehicle Emissions," TRL Published Project Report, 2009.
[31] R.E. Bellman and S.E. Dreyfus, Applied Dynamic Programming, Princeton University Press, 2015.
[32] Arash Mousaei, Nasim Bahari, Guo Mieho, Artificial Neural Networks (ANN) of Proposed Linear Induction Motor with Hybrid Secondary (HLIM) Considering the End Effect, *American Journal of Electrical and Computer Engineering*. Volume 5, Issue 1, June 2021 , pp. 32-39. doi: 10.11648/j.ajece.20210501.15